Comments on
"An implicit-Chebyshev pseudospectral method for the effect of radiation on power-law fluid past a vertical plate immersed in a porous medium", by Nasser S. Elgazery [Communications in Nonlinear Science and Numerical Simulation 13 (2008) 728-744]

Asterios Pantokratoras
Professor of Fluid Mechanics

School of Engineering, Democritus University of Thrace,
67100 Xanthi – Greece
e-mail:apantokr@civil.duth.gr

In the above paper the author treats the boundary layer flow of an electrically conducting, non-Newtonian fluid along a vertical plate embedded in a Darcy-Brinkman-Forchheimer porous medium. The fluid is under the action of a constant transverse magnetic field and the radiation is taken into account in the energy equation. The boundary layer equations are transformed into ordinary ones and subsequently are solved numerically. However, there is a fundamental error in this paper which is presented below:

1. The energy equation (equation 3 in the above paper) used by the author is

$$\frac{\partial T}{\partial t} + u\frac{\partial T}{\partial x} + v\frac{\partial T}{\partial y} = \frac{k_f}{\rho c_p}\frac{\partial^2 T}{\partial y^2} - \frac{1}{\rho c_p}\frac{\partial q_r}{\partial y} \qquad (1)$$

where T is the temperature, t is the time, u and v are the velocities in the x and y directions, $k_f$ is the fluid thermal conductivity, ρ is the fluid density, $c_p$ is the fluid specific heat and $q_r$ is the radiation heat flux. However, the above equation is valid for a clear fluid (without porous medium). In porous media the enegy equation is much different from that in pure fluid flow (Nield and Bejan 1999, page 24, Kaviany 2001, page 148, Bejan 2004, page 576, Sezai 2005, Nield 2007)

$$(\rho c)_m \frac{\partial T}{\partial t} + (\rho c)_f (u \frac{\partial T}{\partial x} + v \frac{\partial T}{\partial y}) = k_m \frac{\partial^2 T}{\partial y^2} - \frac{\partial q_r}{\partial y} \qquad (2)$$

where $(\rho c)_m$ is the heat capacity of the porous medium, $(\rho c)_f$ is the heat capacity of the fluid and $k_m$ is the overall thermal conductivity of porous medium. The relation bettwen the heat capacities is

$$(\rho c)_m = (1-\varepsilon)(\rho c)_s + \varepsilon (\rho c)_f \qquad (3)$$

where $(\rho c)_s$ is the heat capacity of the porous medium material (solid) and $\varepsilon$ is the porosity. The overall thermal conductivity is

$$k_m = (1-\varepsilon)k_s + \varepsilon k_f \qquad (4)$$

where $k_s$ is the thermal conductivity of the porous medium material (solid).
The energy equation used in the above paper is wrong and the presented results also wrong.